\documentclass[pra,twocolumn,nofootinbib,longbibliography,preprintnumbers,amsmath,amssymb]{revtex4}

\usepackage{bm,graphicx,multirow,subfigure,cases}
\usepackage{enumerate}

\newcommand{\numberfield}[1]{\ensuremath{\mathbb{#1}}} 

\newcommand{\cvect}[1] {\textsf{#1}} 

\newcommand{\cmatrix}[1]{\textsf{#1}} 

\newcommand{\set}[1] {\ensuremath{\mathcal{#1}}} 

\newcommand{\mment}[1]{\ensuremath{\mathbf{#1}}} 

\newcommand{\cmment}[1]{\ensuremath{\widetilde{\mathbf{#1}}}} 

\newcommand{\seq}[2]{\ensuremath{[#1,#2]}}   

\newcommand{\lseq}[3]{\ensuremath{[#1,#2,#3]}}   

\newcommand{\llseq}[4]{\ensuremath{[#1,#2,#3,#4]}}   

\newcommand{\toutcome}[1]{\ensuremath{\widetilde{#1}}}

\newcommand{\coutcome}[2]{\ensuremath{(#1\colon\!#2)}}
\newcommand{\coutcomel}[3]{\ensuremath{(#1\colon\!#2\colon\!#3)}}

\newcommand{\rep}[1]{\ensuremath{\hat{#1}}}

\newcommand{\sseq}[1]{\ensuremath{#1}}    

\newcommand{\outcome}[2]{\ensuremath{#1^{(#2)}}} 

\newcommand{\bubble}[2]{\ensuremath(#1,#2)}   

\newcommand{\pll}{\lor}  

\newcommand{\ser}{\ensuremath{\mathop{\bm{\cdot\,}}}}  

\newcommand{\comp}{\times}  

\newcommand{\prob}[1]{\ensuremath{P({#1})}}   

\newcommand{\invseq}[1]{\ensuremath{#1^{-1}}}    

\newcommand{\amp}{z}  

\begin{document}



\title{Derivation of Quantum Theory from Feynman's Rules}

\author{Philip Goyal}
    \email{pgoyal@albany.edu}
    \affiliation{University at Albany~(SUNY), NY, USA}
	\date{\today}

\begin{abstract}
 
Feynman's formulation of quantum theory is remarkable in its combination of formal simplicity and computational power.  However, as a formulation of the abstract structure of quantum theory, it is incomplete as it does not account for most of the fundamental mathematical structure of the standard von Neumann--Dirac formalism such as the unitary evolution of quantum states.  In this paper, we show how to reconstruct the entirety of the finite-dimensional quantum formalism starting from Feynman's rules with the aid of a single new physical postulate, the \emph{no-disturbance postulate}.  This postulate states that a particular class of measurements have no effect on the outcome probabilities of subsequent measurements performed.  We also show how it is possible to derive both the amplitude rule for composite systems of distinguishable subsystems and Dirac's amplitude--action rule, each from a single elementary and natural assumption, by making use of the fact that these assumptions must be consistent with Feynman's rules. 


\end{abstract}

\maketitle

\section{Introduction}

The mathematical formalism of quantum theory has numerous structural features, such as its use of complex numbers, whose physical basis has long been regarded as obscure.  In recent years, there has been a growing interest in deriving these features from compelling physical principles inspired by an informational perspective on physical processes~\cite{Wheeler89, Zeilinger99, Fuchs02,Grinbaum-reconstruction}.  The purpose of such derivation is to better understand the differences between quantum and classical physics, to establish the range of validity of the various parts of the quantum formalism, and to identify physical principles whose validity might extend beyond quantum theory itself.  Substantial progress has now been made, both in deriving much of the quantum formalism from physical principles~\cite{Hardy01a, Clifton-Bub-Halvorson03, DAriano-operational-axioms, Goyal-QT, Reginatto-Schroedinger-derivation, GKS-PRA, Chiribella2011,Masanes-Muller2011,Dakic-Brukner2011}, and in identifying physical principles that account for some of the nonclassical features of quantum physics~\cite{Barrett2006,information-causality}.

Almost without exception, the above-mentioned attempts to understand the quantum formalism have focussed their attention on the standard Dirac--von Neumann formalism.  However, Feynman's formulation of quantum theory provides a strikingly different representation of quantum physics~\cite{Feynman48,FeynmanHibbs65}, and this raises the important question of whether we may be able to gain valuable insights by deriving quantum theory from Feynman's perspective.  

Perhaps the most remarkable feature of Feynman's formulation is its formal simplicity.  The simplicity is achieved by dispensing with the notion of the state of a system and with operators that represent measurements and temporal evolution or symmetry transformations.  Instead, the primary notion is that of a \emph{transition} of a system from one measurement outcome, obtained at some time, to another measurement outcome, obtained at some other time; and a complex-valued \emph{amplitude} is associated with each transition.    Feynman's abstract formalism for individual systems consists of what we shall refer to as \emph{Feynman's rules}~\cite{Feynman48}.  These rules, summarized in Fig.~\ref{fig:feynman-rules}, stipulate how  amplitudes associated with given transitions of a system are combined to yield amplitudes of more complex transitions of that system, and how probabilities are computed from amplitudes.  

As Feynman observed, the rules for combining amplitudes bear a striking resemblance to the rules of probability theory~\cite{Feynman48,FeynmanHibbs65}.  In previous work~\cite{GKS-PRA}, we seized on this observation to \emph{derive} Feynman's rules using a method similar to that used by Cox to derive the rules of probability theory from Boolean algebra~\cite{Cox-PT-paper, Cox61}.  Our derivation provides a particularly clear understanding of  why complex numbers are such a fundamental part of the mathematical structure of quantum theory, and provides a precise understanding of the relationship between Feynman's rules and probability theory~\cite{Goyal-Knuth2011}.

Feynman's rules, however, do not, by themselves, constitute a complete formulation of quantum theory.  Most importantly, they do not account for most of the fundamental mathematical structure of the standard von Neumann--Dirac formalism~\cite{Dirac30, vNeumann55}.   For example, a basic property of the standard formalism is that state evolution is unitary, but this property does not follow from Feynman's rules.  While it is true that Feynman's rules imply unitarity \emph{given} the form of the classical action and Dirac's amplitude--action rule~\cite{Feynman48,JohnsonLapidus2000}, unitarity does not follow as a \emph{direct} consequence of Feynman's rules alone, a problem of which Feynman was aware~\footnote{In Ref.~\cite{Feynman48}, Sec.~11, Feynman states, ``One of the most important characteristics of quantum mechanics is its invariance under unitary transformations\dots. Of course, the present formulation, being equivalent to ordinary formulations, can be mathematically demonstrated to be invariant under these transformations  However, it has not been formulated is such a way that it is \emph{physically} obvious that it is invariant''.}.  Not only is this unsatisfactory on a theoretical level, it is particularly problematic since a corresponding classical action does not always exist for a quantum system~\cite{FeynmanHibbs65}, and, even when one does exist, a mathematically rigorous proof of unitarity on the basis of Feynman's path integral, assuming a general form for the action, is highly nontrivial~\cite{JohnsonLapidus2000}.  

Even more fundamentally, the notion of a quantum state itself does not follow naturally from Feynman's rules.  Contrary to what is asserted in Refs.~\cite{Feynman48, FeynmanHibbs65}, one cannot simply assume that the state of a system consists of the amplitudes to obtain the possible outcomes of a given measurement irrespective of the prior history of the system, for two reasons.  First, one cannot exclude the possibility that the system is entangled with another system, and is therefore in a mixed state rather than a pure state.  Second, before these amplitudes can be declared to constitute the state of the system, one must establish that these amplitudes suffice to compute the outcome probabilities of not just the given measurement but of \emph{any} measurement that could be performed on the system.

Thus, in order to complete the derivation of quantum theory from the Feynman perspective, it is essential to discover what physical ideas are needed in order to derive the standard quantum formalism given Feynman's rules for individual systems.  In this paper, we show that a \emph{single} physical idea, formalized in the \emph{no-disturbance postulate}, suffices.  This postulate states that a particular class of measurements---which we refer to as \emph{trivial} measurements---have no effect on the outcome probabilities of subsequent measurements.  A trivial measurement has a \emph{single} outcome, this outcome having been obtained by coarse-graining over all of the outcomes of an atomic, repeatable measurement~(see Sec.~\ref{sec:experimental} for definitions of these terms).  For example, a Stern-Gerlach measurement with but a single detector which registers all outgoing systems is a trivial measurement.

The no-disturbance postulate formalizes a key difference between classical and quantum physics.  In quantum physics, a trivial measurement is non-disturbing.  However, in classical physics, a trivial measurement is generally disturbing.  For example, consider a trivial Stern-Gerlach measurement with a single coarse-grained outcome obtained by coarse-graining over two atomic outcomes.  From a classical point of view, it is a fact of the matter that the system passed through one of the atomic outcomes, \emph{even though the coarse-grained outcome was not capable of registering this fact}.  As we show in Sec.~\ref{sec:no-disturbance}, this classical inference leads to a change in the outcome probabilities of subsequent measurements.  

One can understand the no-disturbance postulate quite naturally as follows.  From an information-theoretic point of view, it is the gain of information about a quantum system that is ultimately responsible for the disturbance of its state~\cite{Fuchs-Peres1996}.  From this viewpoint, it seems eminently plausible that, conversely, there should exist measurements that provide no useful information which do not disturb the state~\footnote{I am grateful to Paulo Perinotti for suggesting this point of view.}.  Informally, one might say that, if a measurement provides no information, then it need not disturb the state of the system~\footnote{One cannot rule out measurements that provide no useful information and yet still disturb the state.}.  Since a trivial measurement has only one outcome, one gains no information~(in the sense of Shannon information) about which outcome was obtained on learning the outcome of the measurement.  This is similar to ones predicament on learning that a two-headed coin has landed heads.  The no-disturbance postulate asserts that trivial measurements are such non-disturbing, non-informative measurements.   

Since the no-disturbance postulate exposes a fundamental difference between classical and quantum physics, it is an excellent candidate to take as a physical postulate in deriving quantum theory.  Indeed, we have previously employed a special case of this postulate in our derivation of Feynman's rules~\cite{GKS-PRA}, and a very similar idea has more recently been employed in Ref.~\cite{Wehner-Pfister2013} to derive interesting results regarding the state space of general probabilistic theories.

As mentioned above, Feynman's rules concern a given quantum system, so that additional rules must be given if one wishes to treat composite systems of distinguishable or indistinguishable subsystems.  A particularly attractive feature of these rules is their formal simplicity.  In Ref.~\cite{FeynmanHibbs65}, the amplitude rules for such composite systems are simply postulated,  presumably having been extracted from the standard formalism.   In this paper, we show that the rule for composite systems of distinguishable subsystems can in fact be \emph{derived} from a simple \emph{composition postulate} on the condition that  Feynman's rules for individual systems are valid.  The composition postulate simply posits that the amplitude of a transition of a composite system consisting of two noninteracting subsystems is a continuous function of the amplitudes of the respective subsystem transitions.    Remarkably, the rule for assigning amplitudes to composite systems is uniquely determined by the requirement that the amplitude assignment to the composite system be consistent with Feynman's rules for assigning amplitudes to individual systems.  This rule immediately gives rise to the tensor product rule in the standard quantum formalism.   The derivation of the symmetrization postulate, which is needed to describe composite systems consisting of indistinguishable subsystems, is detailed elsewhere~\cite{Goyal2013a}.  

Finally, a striking feature of Feynman's formulation is its remarkably direct connection to the Lagrangian formulation of classical physics.  More precisely, when a series of transitions of a quantum system in configuration space is well approximated by a continuous trajectory of the corresponding classical system, Dirac's amplitude--action rule associates the amplitude~$e^{iS/\hbar}$ to the series of transitions, where~$S$ is the action of the corresponding classical trajectory~\cite{Dirac33}.  It is this rule from which much of the computational power of Feynman's rules derives, allowing, for example, the derivation of the Schroedinger equation~\cite{Feynman48} and quantum electrodynamics~\cite{Feynman49}.   Dirac obtained this rule on the basis of a detailed analogy between transformations in quantum theory and contact transformations in the Lagrangian formulation of classical mechanics.  However, given the fundamental importance of this classical--quantum connection, a simpler and more direct derivation is desirable. Here, using only  elementary properties of the action, we provide a simple and direct derivation of this rule on the simple assumption that the amplitude of a path in configuration space is a continuous function of its classical action. 

The work described here significantly improves upon previous attempts to address many of the above-mentioned issues.  In particular, a previous attempt to derive unitarity from Feynman's rules makes appeal to the Hilbert space norm, which itself must be independently justified~\cite{Caticha99b}.  In contrast, our derivation rests entirely on the no-disturbance postulate.  Similarly, two previous derivations of the rule for composite systems of distinguishable subsystems implicitly assume that the functional relationships involved are complex-analytic, an assumption which significantly detracts from the physical transparency of the derivations~\cite{Tikochinsky-identical, Caticha98b}.  We are able to avoid such assumptions.

The results given here have implications for many issues which have been raised in the literature.  For example, a recurrent question is whether linear temporal evolution can be replaced with nonlinear temporal evolution~\cite{Pearle1976, Birula1976, Shimony1979, Weinberg89a,Weinberg89b}.  Such replacement has a variety of motivations, such as the desire to incorporate the quantum measurement process into the usual temporal evolution of a quantum state, or to solve the black hole information paradox~\cite{Giddings1995}.  Such work has led to attempts to explore whether, taking certain other parts of the quantum formalism as a given, one can derive linearity or unitarity from a physical principle such as the requirement that there is no instantaneous signaling between separated subsystems~\cite{Simon2001,Ferrero2004,Ferrero2006}.   Our earlier derivation of Feynman's rules, together with the present derivation of unitarity from the no-disturbance postulate, shows that the linearity and unitarity of temporal evolution is very basic to the structure of quantum theory, and cannot be replaced with nonlinear deterministic evolution, even in a manner that is barely experimentally perceptible, without undermining the entire edifice.  Furthermore, since our derivation of unitarity depends only on no-disturbance, and not on any postulate that refers to the behavior~(such as no instantaneous signaling) of composite systems, our result shows that such behavior, contrary to that which is suggested by some previous work~\cite{Simon2001,Ferrero2004,Ferrero2006}, is not necessary to understand unitarity.


The remainder of the paper is organized as follows.  In Sec.~\ref{sec:background}, we summarize the experimental framework and notation presented in Ref.~\cite{GKS-PRA}, extend the framework to deal with composite systems, and summarize Feynman's rules in operational language.  In Sec.~\ref{sec:state-formulation}, we formulate the no-disturbance postulate, and use this to systematically introduce the notion of a quantum state, show that states evolve unitarily, and show that repeatable measurements are represented by hermitian operators.   In Sec.~\ref{sec:composite-systems}, we introduce a composition operator, formulate its symmetry relations, and show, via the composition postulate, that these lead to the amplitude rule for composite systems.  Finally, in Sec.~\ref{sec:amplitude-action}, we derive Dirac's amplitude--action rule.  We conclude in Sec.~\ref{sec:discussion} with a discussion of the results and their broader implications.

\section{Background and Notation}
\label{sec:background}

\subsection{Experimental Framework}
\label{sec:experimental} 

\begin{figure}
\includegraphics[width=3.5in]{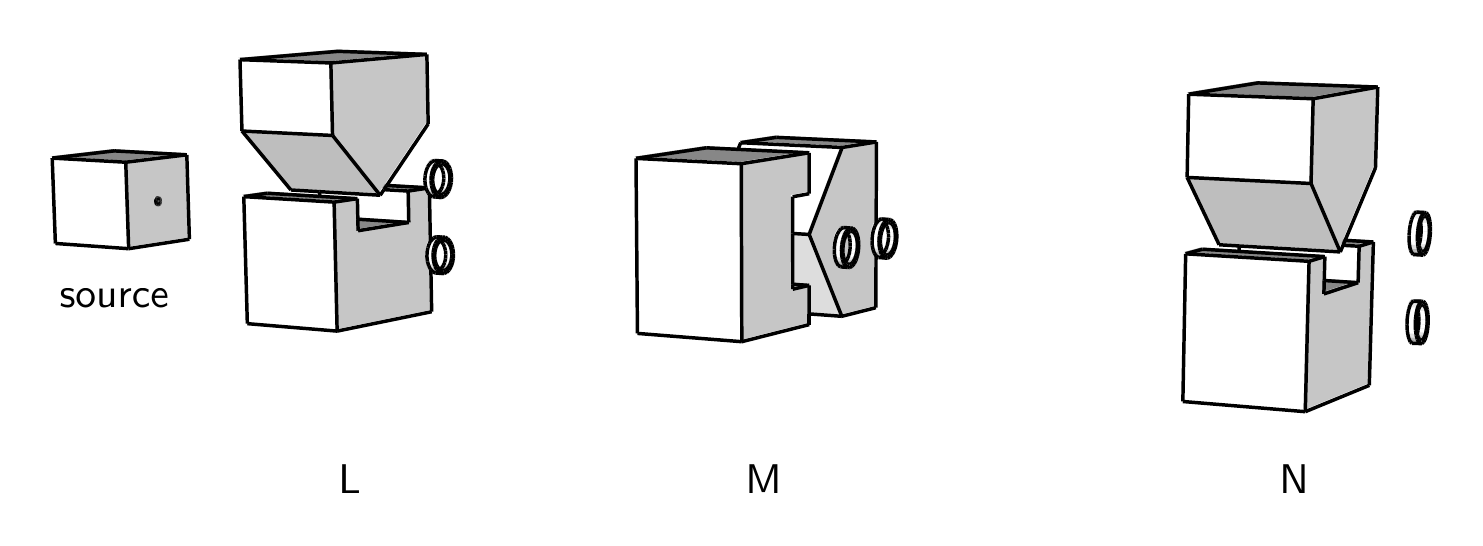}
\caption{\label{fig:SG-example} Schematic representation of a Stern-Gerlach experiment performed on silver atoms.  A silver atom from a source~(an evaporator) is subject to a sequence of measurements, each of which yields one of two possible outcomes registered by non-absorbing wire-loop detectors.  Between measurements, the atoms interact with a uniform magnetic field.  A run of the experiment yields outcomes~$\ell, m, n$ of the measurements~$\mment{L}, \mment{M}, \mment{N}$ performed at times~$t_1, t_2, t_3$, respectively.  The probability distribution over the outcomes of~$\mment{M}$ given an outcome of~$\mment{L}$ is observed to be independent of any interactions the system had prior to~$\mment{L}$, a property to which we refer as closure. }
\end{figure}

An experimental set-up is defined by specifying a source of physical systems, a sequence of measurements to be performed on a system in each run of the experiment, and the interactions between the system and its environment which occur between the measurements~(see example in Fig.~\ref{fig:SG-example}).   In a run of an experiment, a physical system from the source passes through a sequence of measurements~$\mment{L}, \mment{M}, \mment{N}, \dots$, which respectively yield outcomes~$\ell, m, n, \dots$ at times~$t_1, t_2, t_3, \dots$.  These outcomes are summarized in the \emph{measurement sequence} $\llseq{\ell}{m}{n}{\dots}$.  In between these measurements, the system may undergo interactions with the environment.   We shall label the possible outcomes of a measurement~$\mment{M}$ as~$m, m', m'', m''', \dots$ as far as is needed in each case.

Over many runs of the experiment, the experimenter will observe the frequencies of the various possible measurement sequences, from which the experimenter can estimate the probability associated with each sequence.   The probability~$\prob{A}$ associated with sequence~$\sseq{A} = \llseq{\ell}{m}{n}{\dots}$ is defined as the probability of obtaining outcomes~$m, n, \dots$ conditional upon obtaining~$\ell$,
\begin{equation} \label{eqn:def-of-probability-of-sequence}
\prob{A} = \Pr(m, n, \dots \,|\, \ell).
\end{equation}

A particular outcome of a measurement is either \emph{atomic} or \emph{coarse-grained}.   An atomic outcome is one that cannot be more finely divided in the sense that the detector whose output corresponds to the outcome cannot be sub-divided into smaller detectors whose outputs correspond to two or more outcomes.  An example of atomic outcomes are the two possible outcomes of a Stern-Gerlach measurement performed on a silver atom.  A coarse-grained outcome is one that does not differentiate between two or more outcomes, an example being a Stern-Gerlach measurement where a detector's field of sensitivity encompasses the fields of sensitivity of two detectors, each of which corresponds to a different atomic outcome.    Abstractly, if a measurement has an outcome which is a coarse-graining of the outcomes labeled~$m$ and~$m'$ of measurement~$\mment{M}$, the outcome is labeled~$\bubble{m}{m'}$, and this notational convention naturally extends to coarse-graining of more than two atomic outcomes.    In general, if all of the possible outcomes of a measurement are atomic, we shall call the measurement itself atomic.  Otherwise, we say it is a coarse-grained measurement.  A coarse-grained measurement with but a single outcome is called a \emph{trivial} measurement as its outcome provides us with no more information than the fact that the measurement has detected a system at a particular time.

As explained in Refs.~\cite{Goyal-QT2c, GKS-PRA}, the measurements and interactions which can be employed in a given experiment must satisfy certain conditions if they are to lead to a well-defined theoretical model. The measurements that are employed in an experimental set-up must be repeatable, come from the same \emph{measurement set},~$\set{M}$, or be coarsened versions of measurements drawn from this set; and the first measurement in each experiment must be atomic.  These conditions ensure that (i)~all the measurements are probing the same aspect~(say, the spin behavior) of the system, and (ii)~the outcome probabilities of all measurements except the first are independent of the history of the system prior to the experiment, a property we refer to as \emph{closure}.    Similarly, interactions that occur in the period of time between measurements are selected from a set,~$\set{I},$ of possible interactions.  These interactions preserve closure when they act on the given system between any two measurements in~$\set{M}$.  For the operational definitions of sets~$\set{M}$ and~$\set{I}$, and further discussion of the above conditions, we refer the reader to Ref.~\cite{GKS-PRA}.


\subsubsection*{Composite Systems}

Suppose that two distinguishable physical systems,~$S_1$ and~$S_2$, simultaneously undergo experiments.  Operationally, distinguishability means that there is some measurement which can be performed on each system before and after an experiment which determines the identity of the system passing through the experiment.    In particular, suppose that system~$S_1$ undergoes an experiment where measurements~$\mment{L}_1, \mment{M}_1, \mment{N}_1$ are performed at successive times~$t_1, t_2, t_3$, while the other system,~$S_2$, undergoes a separate experiment where measurements~$\mment{L}_2, \mment{M}_2, \mment{N}_2$ are performed at these times.   

\begin{figure}
\centering
\subfigure[]
				{
				\includegraphics[width=3.5in]{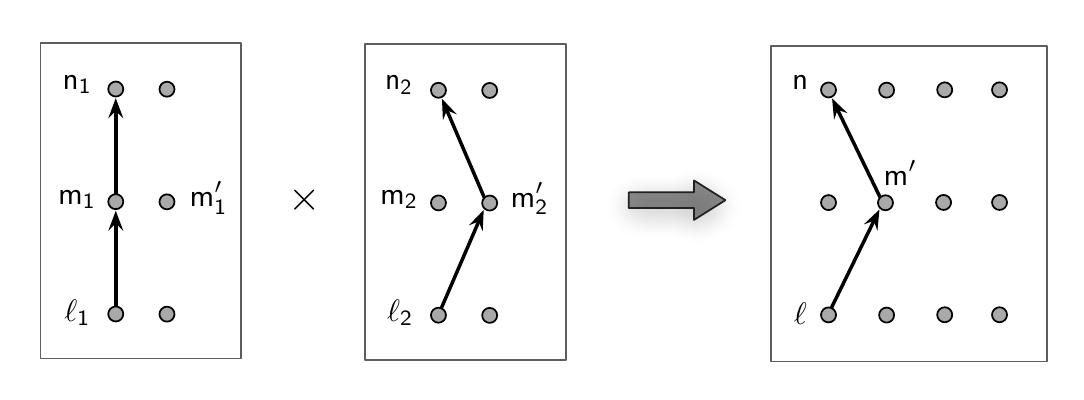}
				\label{fig:comp-example1}
				}
\subfigure[]
				{
				\includegraphics[width=3.5in]{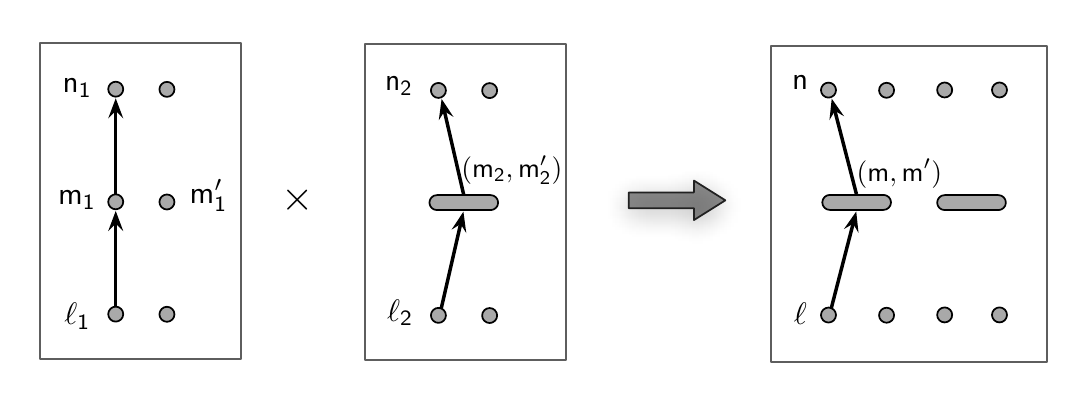}
				\label{fig:comp-example2}
				}
\caption[]{\label{fig:comp-example} Composition of sequences belonging to two separate systems.  In both examples, the first system undergoes measurements~$\mment{L}_1, \mment{M}_1, \mment{N}_1$.  The second system undergoes~$\mment{L}_2, \mment{M}_2, \mment{N}_2$ in~(a) and~$\mment{L}_2, \cmment{M}_2, \mment{N}_2$ in~(b).   For concreteness, the atomic measurements are assumed to have two possible outcomes, with~$m_1, m_1'$ labelling the possible outcomes of~$\mment{M}_1$, and so on.   In~(a), the first system yields sequence~$A = \lseq{\ell_1}{m_1}{n_1}$, and the second system yields~$B = \lseq{\ell_2}{m_2'}{n_2}$, which are composed~($\times$) to yield sequence~$C  = \lseq{\ell}{m'}{n} = \lseq{\coutcome{\ell_1}{\ell_2}}{\coutcome{m_1}{m_2'}}{\coutcome{n_1}{n_2}}$.  In~(b), the first system yields~$A = \lseq{\ell_1}{m_1}{n_1}$, and the second system yields~$B = \lseq{\ell_2}{\bubble{m_2}{m_2'}}{n_2}$, which compose to yield~$C  = \lseq{\ell}{\bubble{m}{m'}}{n} = \lseq{\coutcome{\ell_1}{\ell_2}}{\bubble{\coutcome{m_1}{m_2}}{\coutcome{m_1}{m_2'}}}{\coutcome{n_1}{n_2}}$.}
\end{figure}

These two experiments can equivalently be viewed as a \emph{single} experiment performed on a system,~$S$, undergoing a sequence of measurements~$\mment{L}, \mment{M}, \mment{N}$, where each of these measurements is viewed as a composite of the corresponding indexed measurements; for example~$\mment{L}$ is viewed as a composite of~$\mment{L}_1$ and~$\mment{L}_2$.  We then say that~$S$ is a \emph{composite} system consisting of subsystems~$S_1$ and~$S_2$.  If, say, measurement~$\mment{L}_1$ yields outcome~$\ell_1$, and measurement~$\mment{L}_2$ yields~$\ell_2$, this can be described as measurement~$\mment{L}$ yielding an outcome~$\ell$ which we symbolize as~$\coutcome{\ell_1}{\ell_2}$.  The process of composition is illustrated in Fig.~\ref{fig:comp-example}.

In order that the experiments on~$S_1$ and on~$S_2$ separately satisfy the experimental conditions stated above, the systems cannot be interacting with one another while the experiments are in progress.  If the systems are interacting with one another, then only the two experiments viewed as a whole---as an experiment on~$S$---can satisfy these conditions.



%
%


\subsection{Operationalization of Feynman's paths and Feynman's rules}

\label{sec:operational-logic}

Consider an experimental set-up in which a physical system is subject to successive measurements~$\mment{L}, \mment{M}, \mment{N}$ at successive times~$t_1, t_2, t_3$, with there possibly being interactions with the system in the intervals between those measurements.  Here and subsequently, we assume that the measurements and interactions in any such set-up are selected according to the constraints described above.  We summarize the outcomes obtained in a given run of the experiment as the \emph{sequence}~$\sseq{C} = \lseq{\ell}{m}{n}$.  This is the operational counterpart to a Feynman `path'.

We now wish to formalize the idea that set-ups are interrelated in particular ways.  In Ref.~\cite{GKS-PRA}, we considered two such relationships.  First, the above set-up could be viewed as a \emph{series concatenation} of two experiments, the first in which measurements~$\mment{L}$ and~$\mment{M}$ occur at times~$t_1$ and~$t_2$, yielding the sequence~$\sseq{A} = \seq{\ell}{m}$, and the second in which measurements~$\mment{M}$ and~$\mment{N}$ occur at times~$t_2$ and~$t_3$, yielding~$\sseq{B} = \seq{m}{n}$.  In order to ensure that experimental closure is satisfied in the second experiment, measurement~$\mment{M}$ must be atomic.  Formally, we express this concatenation as
\begin{equation}
\sseq{C} = \sseq{A} \ser \sseq{B},
\end{equation}
where~$\ser$ is the \emph{series} combination operator.  More generally, the series operator can be used to concatenate two sequences provided their initial and final measurements are atomic, and the final measurement and outcome of the first sequence is the same as the initial measurement and outcome of the second sequence.   

Second, one can consider a set-up which is identical to the one above, except that outcomes~$m$ and~$m'$ of~$\mment{M}$ have been coarse-grained, so that one obtains the sequence~$\sseq{E} = \lseq{\ell}{\bubble{m}{m'}}{n}$.  Formally, we express the relationship of this sequence to the sequences~$\sseq{C} = \lseq{\ell}{m}{n}$ and~$\sseq{D} = \lseq{\ell}{m'}{n}$ as
\begin{equation}
\sseq{E} = \sseq{C} \pll \sseq{D},
\end{equation}
where~$\pll$ is the \emph{parallel} combination operator.  More generally, the parallel operator can combine any two sequences which are identical except for differing in the outcome of a single measurement in the set-up, provided that this measurement is not the initial or final measurement.

%
%

\subsubsection*{Feynman's Rules in Operational Form}

From the above definitions, it follows that the operators~$\ser$ and~$\pll$ 
satisfy several symmetry relations, to which we collectively refer to as an \emph{experimental logic}:
\begin{align} 
A \pll B   &= B \pll A										\label{eqn:parallel-commutativity}  \\
(A \pll B) \pll C    &= A  \pll (B \pll C) 				\label{eqn:parallel-associativity}  \\
(A \ser B) \ser C    &= A  \ser \, (B \ser C)			\label{eqn:series-associativity}  \\
 (A \pll B) \ser C &= (A \ser C) \pll (B \ser C)		\label{eqn:right-distributivity} \\
C \ser\, (A \pll B) &= (C \ser {A}) \pll (C \ser B)	\label{eqn:left-distributivity}
\end{align}

\begin{figure}[!h]
\centering
\subfigure[\,\emph{Sum rule:}~$\amp(A\pll B) = \amp(A) + \amp(B)$.]
				{
				\includegraphics[width=2.5in]{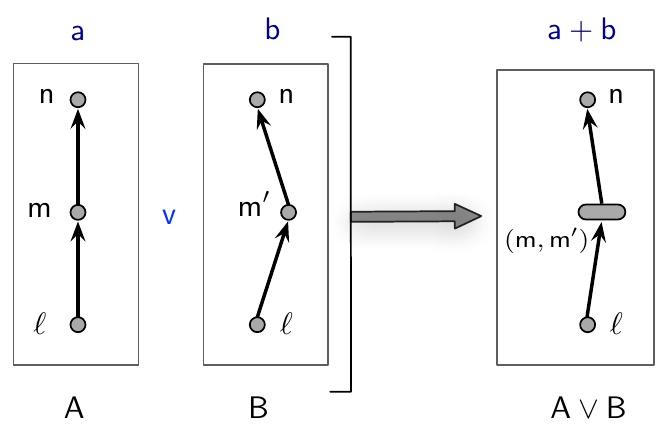}
				\label{fig:sum}
				}
\subfigure[\,\emph{Product rule:}~$\amp(A\ser B) = \amp(A)\cdot\amp(B)$.]
				{
				\includegraphics[width=2.5in]{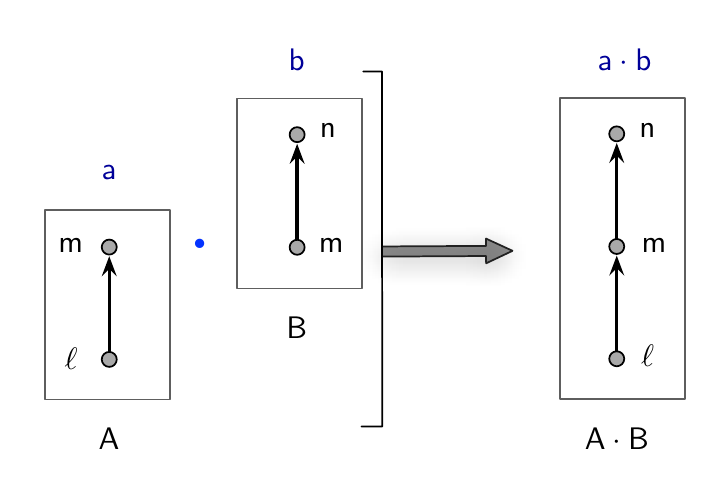}
				\label{fig:product}
				}
\subfigure[\,\emph{Probability rule:}~$\Pr(m, n\,|\,\ell) = |a|^2$.]
				{
				\includegraphics[width=2.5in]{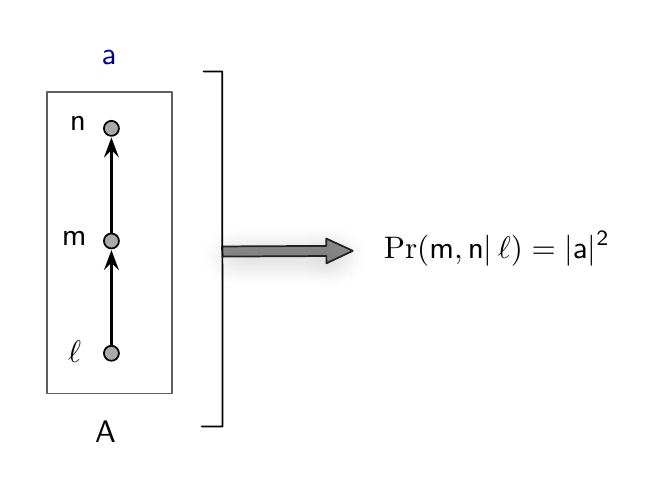}
				\label{fig:probability-rule}
				}
\caption[]{\label{fig:feynman-rules} Feynman's rules for individual systems, expressed in operational terms.  In each case, the sequence names are denoted~$A, B, \dots$, while their amplitudes are denoted~$a, b, \dots$.}
\end{figure}

In Ref.~\cite{GKS-PRA}, it is shown that Feynman's rules are the unique pair-valued representation of this logic consistent with a few additional assumptions.  Writing~$\amp({X})$ for the complex-valued \emph{amplitude} that represents sequence~$\sseq{X}$, one finds~(see Fig.~\ref{fig:feynman-rules}):
\begin{align}
\amp(\sseq{A} \pll \sseq{B}) 	&= \amp(\sseq{A}) + \amp(\sseq{B}) 	\tag{amplitude sum rule}\\
\amp(\sseq{A} \ser \sseq{B}) 	&= \amp(\sseq{A}) \cdot \amp(\sseq{B}) \tag{amplitude product rule}\\
P(\sseq{A}) 		&= \left| \amp(\sseq{A}) \right|^2.		\tag{probability rule}
\end{align}
These are Feynman's rules for measurements on individual quantum systems.  

\section{State Formulation of Quantum Theory}
\label{sec:state-formulation}

In the standard, von Neumann--Dirac formulation of quantum theory, one describes a system by specifying its state at a particular time.  Temporal evolution of the system is then represented by a unitary operator, and repeatable measurements made on the system are represented by Hermitian operators.  When the states of the subsystems of a composite system are given, the state of the composite system is the tensor product of the subsystem states.  In this section, starting from Feynman's rules and the composite systems rule~(derived in Sec.~\ref{sec:composite-systems}), we derive these features with the aid of the no-disturbance postulate.

\subsection{No-disturbance postulate}
\label{sec:no-disturbance}

\begin{figure}
\centering
\includegraphics[width=2.75in]{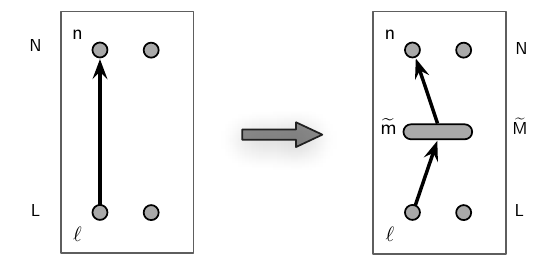}
\caption{\label{fig:no-disturbance} \emph{No-disturbance postulate.}  \emph{Left:} A system undergoes measurement~$\mment{L}$ at time~$t$, followed by~$\mment{N}$ at time~$t'$.  For illustration, each measurement has two possible outcomes.  The sequence of outcomes~$\seq{\ell}{n}$ has associated probability~$\Pr(n\,|\,\ell)$. \emph{Right:}  Trivial measurement~$\cmment{M}$, with single outcome~$\toutcome{m}$, occurs between~$\mment{L}$ and~$\mment{N}$.  By the no-disturbance postulate,~$\cmment{M}$ has no effect on the probability of outcome~$n$ given~$\ell$.}
\end{figure}

The no-disturbance postulate asserts that a trivial measurement~(as defined in Sec.~\ref{sec:experimental}) has no effect on the outcome probabilities of subsequent measurements.   For example, in the arrangement shown in Fig.~\ref{fig:no-disturbance}, if the trivial measurement~$\cmment{M}$, with single outcome~$\toutcome{m} = \bubble{m}{m'}$, is inserted between measurements~$\mment{L}$ and~$\mment{N}$, the probability of outcome~$n$ given~$\ell$ is unaffected.  
That is,
\begin{equation}
\Pr(n\,|\,\ell; \cmment{M}) = \Pr(n\,|\,\ell),
\end{equation}
where~$\cmment{M}$ in the conditional on the left-hand side indicates that the arrangement containing~$\cmment{M}$ is the one under consideration.

%
%
%
%
%
%

\begin{figure}
\centering
\includegraphics[width=2.75in]{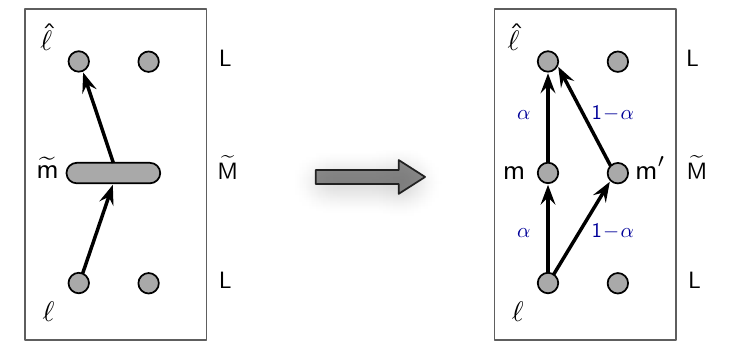}
\caption{\label{fig:disturbance} \emph{Disturbance of repeatability.}  \emph{Left:} A system undergoes measurement~$\mment{L}$ at time~$t$, and again immediately afterwards at~$t'$, with trivial measurement~$\cmment{M}$ in between.  Since measurement~$\mment{L}$ is a repeatable measurement, the no-disturbance postulate implies that it will yield the same outcome at~$t'$ as at~$t$, even though~$\cmment{M}$ is present.   \emph{Right:}  From a classical point of view, the occurrence of~$\toutcome{m} = \bubble{m}{m'}$ implies that either outcome~$m$ or~$m'$ occurred, but was not observed.  The transition probabilities are as indicated, assuming that transition probabilities are symmetric.  From these probabilities, it follows that repeatability is disturbed by~$\cmment{M}$ unless~$\alpha$ is~$0$ or~$1$.}
\end{figure}

The no-disturbance postulate can be regarded as capturing the essential departure of quantum physics from the mode of thinking embodied in classical physics.  The key point is that, from the classical point of view, if outcome~$\toutcome{m}$ is obtained, one would assert that it is a fact of the matter that the system went either through the field of sensitivity of the detector corresponding to outcome~$m$ or through that corresponding to~$m'$, even though neither was, in fact, observed.   To see the consequences of this assertion, let us consider the special case where~$\mment{L}$ is repeated at time~$t'$,  where~$t'$ is immediately after~$t$ so that the system undergoes no appreciable temporal evolution in the interim~(see Fig.~\ref{fig:disturbance}). 
For clarity, we denote outcome~$\ell$ of measurement~$\mment{L}$ at~$t'$ by~$\rep{\ell}$.  Now, according to the classical assertion,
\begin{align*}
\Pr(\rep{\ell}\,|\,\ell; \cmment{M}) 	&= \Pr(\rep{\ell}, m\,|\,\ell) + \Pr(\rep{\ell}, m'\,|\,\ell) \\
					&= \Pr(m\,|\,\ell) \, \Pr(\rep{\ell}\,|\, m, \ell) + \Pr(m'\,|\,\ell) \, \Pr(\rep{\ell}\,|\, m', \ell) \\
					&= \Pr(m\,|\,\ell) \, \Pr(\rep{\ell}\,|\, m) + \Pr(m'\,|\,\ell) \, \Pr(\rep{\ell}\,|\, m'),
\end{align*}
where we have used the sum and product rules of probability theory in the first two lines, and closure in the third.  If we now assume that transition probabilities are symmetric~(an assumption that is independently well-supported by experiment), then~$\Pr(\rep{\ell} \, | \, m) = \Pr(m \,|\, \rep{\ell})$ and~$\Pr(\rep{\ell} \, | \, m') = \Pr(m' \,|\, \rep{\ell})$.  Setting~$\alpha = \Pr(m\,|\,\ell)$ and noting that~$\Pr(m\,|\,\ell) + \Pr(m'\,|\,\ell) = 1$,
\begin{align*}
\Pr(\rep{\ell}\,|\,\ell; \cmment{M})	&= \Pr(m\,|\,\ell) \,  \Pr(m \,|\, \rep{\ell}) + \Pr(m'\,|\,\ell) \, \Pr(m' \,|\, \rep{\ell}) \\
													&= \alpha^2 + (1 - \alpha)^2.
\end{align*}
Since~$\mment{L}$ is a repeatable measurement,~$\Pr(\rep{\ell}\,|\,\ell; \cmment{M})$ should be unity.  But, for this to be possible, $\alpha^2 + (1 - \alpha)^2 = 1$, which cannot hold true unless~$\alpha$ is~$0$ or~$1$.  Therefore, repeatability cannot be preserved by the insertion of~$\cmment{M}$ except in the special case where one of the outcomes~$m$ or~$m'$ is certain to occur.   That is, on the classical assertion that the occurrence of outcome~$\toutcome{m}$ implies that either outcome~$m$ or~$m'$ in fact occurs, insertion of~$\cmment{M}$ will, in general, disturb repeatability of~$\mment{L}$.   Conversely, if the no-disturbance postulate is true, one must conclude that the classical assertion is, in general, false.  

\subsection{Quantum States}

We operationally define the mathematical representation of the physical state of a system at any given time as that mathematical object which enables one to compute the outcome probabilities of any measurement~(chosen from a given measurement set~$\set{M}$) performed upon the system at that time.  


First, suppose that a system is prepared at time~$t$ using measurement~$\mment{L}$, and that measurement~$\mment{M}$ is subsequently performed upon it at time~$t'$.  Here, and subsequently, we assume that all measurements belong to the same measurement set, and each have~$N$ possible outcomes.  We label the~$j$th outcome of measurement~$M$ as~$\outcome{m}{j}$, where~$j \in \{1,2,\dots, N\}$, and the outcomes of other measurements similarly.  

Suppose that measurement~$\mment{L}$ yields outcome~$\outcome{\ell}{i}$.  In order to compute the transition probabilities~$\Pr(\outcome{m}{j}\,|\,\outcome{\ell}{i})$ for every~$j$, it suffices to know the amplitude vector~$\cvect{v} = (v_1,  \dots, v_N)$ whose~$j$th component is the amplitude of the sequence~$\seq{\outcome{\ell}{i}}{\outcome{m}{j}}$.  Then, $\Pr(\outcome{m}{j}\,|\,\outcome{\ell}{i}) = |v_j|^2$.  Since the outcomes of~$\mment{M}$ are mutually exclusive and exhaustive,~$|\cvect{v}|^2 = \sum_j |v_j|^2 = 1$.  Insofar as calculating the outcome probabilities of~$\mment{M}$ performed at~$t'$, the object~$\cvect{v}$ suffices.

\begin{figure}
\centering
\includegraphics[width=3.25in]{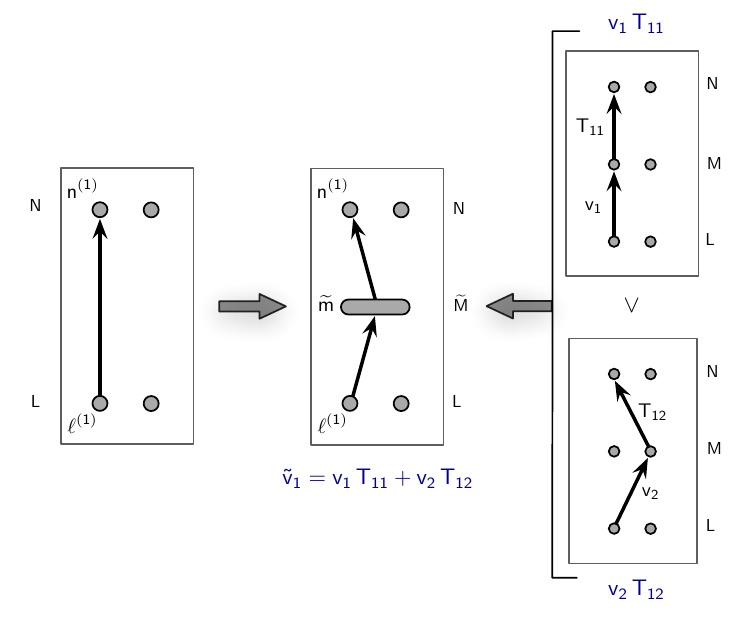}
\caption{\label{fig:introducing-states} \emph{Left:} A system undergoes measurement~$\mment{L}$ at time~$t$, followed by~$\mment{N}$ at time~$t'$, yielding sequence~$\seq{\outcome{\ell}{1}}{\outcome{n}{1}}$.  \emph{Middle:} If trivial measurement~$\cmment{M}$ occurs immediately prior to~$\mment{N}$, then, by the no-disturbance postulate, it has no effect on the outcome probabilities of~$\mment{N}$.  Hence, the probability~$\Pr(\outcome{n}{1}\,|\,\outcome{\ell}{1}) = |\tilde{v}_1|^2$. \emph{Right:} From the amplitude sum rule,~$\tilde{v}_1  = v_1T_{11} + v_2 T_{12} = (\cmatrix{T}\cvect{v})_1.$}
\end{figure}

Second, suppose that, at time~$t'$, instead of~$\mment{M}$, we wish to perform measurement~$\mment{N}$, and to compute its outcome probabilities.  To do so, we now make use of the no-disturbance postulate, according to which we can insert the trivial form of measurement~$\mment{M}$, which we denote~$\cmment{M}$, prior to measurement~$\mment{N}$, without changing the outcome probabilities of~$\mment{N}$.  We insert~$\cmment{M}$ immediately prior to~$\mment{N}$ in order that the system undergo no appreciable temporal evolution between~$\cmment{M}$ and~$\mment{N}$~(see Fig.~\ref{fig:introducing-states}).  
We can now compute the outcome probabilities of~$\mment{N}$ in the modified arrangement instead.  Now, in this arrangement, the sequence~$\lseq{\outcome{\ell}{i}}{\widetilde{m}}{\outcome{n}{k}}$, where~$\widetilde{m} \equiv (\outcome{m}{1}, \dots,\outcome{m}{N})$, can be decomposed as
\begin{align}
\lseq{\outcome{\ell}{i}}{\widetilde{m}}{\outcome{n}{k}} &= \bigvee_j \,\lseq{\outcome{\ell}{i}}{\outcome{m}{j}}{\outcome{n}{k}} \\
																			&= \bigvee_j \,\seq{\outcome{\ell}{i}}{\outcome{m}{j}} \ser \seq{\outcome{m}{j}}{\outcome{n}{k}}.
\end{align}
Hence, given the amplitudes~$T_{kj}$ of the sequences~$\seq{\outcome{m}{j}}{\outcome{n}{k}}$, the amplitude sum and product rules imply that the amplitude~$\tilde{v}_k$ of sequence~$\lseq{\outcome{\ell}{i}}{\widetilde{m}}{\outcome{n}{k}}$ is
\begin{equation}
\tilde{v}_k =  \sum_{j} v_j T_{kj} = (\cmatrix{T}\cvect{v})_k,
\end{equation}
where~$\cmatrix{T}$ is a matrix with components~$T_{kj}$.  Since the system undergoes no appreciable temporal evolution between measurement~$\cmment{M}$ and~$\mment{N}$, the matrix~$\cmatrix{T}$ captures precisely the relationship between~$\mment{M}$ and~$\mment{N}$.  We shall refer to it as the transformation matrix from~$\mment{M}$ to~$\mment{N}$.  Hence, the transition probability
\begin{equation}
\Pr(\widetilde{m}, \outcome{n}{k}\,|\,\outcome{\ell}{i}) = |(\cmatrix{T}\cvect{v})_k|^2.
\end{equation}
Using the product rule of probability theory,
\begin{equation}
\Pr(\widetilde{m}, \outcome{n}{k}\,|\,\outcome{\ell}{i}) = \Pr(\outcome{n}{k}\,|\,\outcome{\ell}{i}) \, \Pr(\widetilde{m} \,|\,\outcome{n}{k}, \outcome{\ell}{i}),
\end{equation}
and noting that~$ \Pr(\widetilde{m} \,|\,\outcome{n}{k}, \outcome{\ell}{i}) = 1$, we obtain
\begin{equation} \label{eqn:l-n-probability}
\Pr(\outcome{n}{k}\,|\,\outcome{\ell}{i}) = |(\cmatrix{T}\cvect{v})_k|^2.
\end{equation}
This statement holds for the modified experiment in which~$\cmment{M}$ occurs.  But, by the non-disturbance postulate, it also holds true for the original experiment.  

Thus, the object~$\cvect{v}$, which is specified with respect to~$\mment{M}$, not only allows one to compute the outcome probabilities of~$\mment{M}$, but \emph{also} to compute the outcome probabilities of any other measurement,~$\mment{N}$, provided one is given the transformation matrix,~$\cmatrix{T}$, from~$\mment{M}$ to~$\mment{N}$.  Therefore, from the operational point of view stated earlier, \emph{$\cvect{v}$ represents the state of the system at time~$t'$.}  



\subsection{Representation of Measurements}

Since the outcomes of~$\mment{N}$ are mutually exclusive and exhaustive, Eq.~\eqref{eqn:l-n-probability} becomes
\begin{equation}
\sum_k \Pr(\outcome{n}{k}\,|\,\outcome{\ell}{i}) = \sum_k |(\cmatrix{T}\cvect{v})_k|^2 = 1,
\end{equation}
which implies that~$|\cmatrix{T}\cvect{v}|^2 = 1$.  But~$\cvect{v}$ can be freely varied by varying the initial measurement~$\mment{L}$, its outcome~$\outcome{\ell}{i}$, and the interaction with the system in the interval~$[t, t']$.  Therefore, the transformation matrix,~$\cmatrix{T}$, which connects~$\mment{M}$ to~$\mment{N}$, is unitary.

To determine the states that can be prepared by measurement~$\mment{N}$, we use the fact that, since~$\mment{N}$ is repeatable, if a system is prepared at time~$t$ using measurement~$\mment{N}$ with outcome~$\outcome{n}{q}$, the same outcome is obtained when the measurement is immediately repeated.  Therefore, using Eq.~\eqref{eqn:l-n-probability}, the state,~$\cvect{u}_q$, that is prepared must be such that
\begin{equation}
\Pr(\outcome{n}{k} \, | \, \outcome{n}{q}) = |(\cmatrix{T}\cvect{u}_q)_k|^2 = \delta_{qk}, 
\end{equation}
which implies that~$\cvect{u}_q = (T_{q1}, \dots, T_{qN})^\dagger$ up to a predictively irrelevant overall phase.   In terms of the~$\cvect{u}_q$, one can write~$\tilde{v}_k= \cvect{u}_k^\dagger\cvect{v}$, so that Eq.~\eqref{eqn:l-n-probability} becomes
\begin{equation} \label{eqn:Born}
\Pr(\outcome{n}{k}\,|\,\outcome{\ell}{i}) = |\cvect{u}_k^\dagger\cvect{v}|^2,
\end{equation}
which is the Born rule with~$\cvect{u}_k$ and~$\cvect{v}$ specified with respect to~$\mment{M}$.  Therefore, measurement~$\mment{N}$ can be characterized in terms of the~$\cvect{u}_q$, which form an orthonormal basis of~$\numberfield{C}^N$.  Alternatively, as is more conventional, we can represent~$\mment{N}$ in terms of the Hermitian matrix~$\cmatrix{N} = \sum_q a_q \cvect{u}_q \cvect{u}_q^\dagger$, where~$a_q$ is the value associated with outcome~$\outcome{n}{q}$.

In the special case where measurement~$\mment{N}$ is the same as~$\mment{M}$, it follows from the repeatability of measurements that~$\Pr(\outcome{m}{k}\,|\,\outcome{m}{j}) = \delta_{jk}$.  Therefore the transformation matrix~$\cmatrix{T}'$ that relates~$\mment{M}$ to itself has the property that~$T'_{kj} = \delta_{kj} \,e^{i\phi_k}$, where the~$\phi_{k}$ are phases.  Now, the states~$\cvect{u}'_q = (T'_{q1}, \dots, T'_{qN})^\dagger$ prepared by~$\mment{M}$ are predictively relevant only via Eq.~\eqref{eqn:Born}, whose result is insensitive to the values of the~$\phi_k$.  Therefore, without loss of generality, the~$\phi_{k}$ can all be set to zero, so that~$\cmatrix{T}'$ reduces to the identity matrix~$\cmatrix{I}$.  Hence, measurement~$\mment{M}$ is represented by a diagonal Hermitian matrix,~$\cmatrix{M}$.

\subsection{Relationship between Representations}

Thus far, we have specified the state of the system,~$\cvect{v}$, and the states~$\cvect{u}_k$ that are prepared by measurement~$\mment{N}$, with respect to measurement~$\mment{M}$.    Suppose that we were instead to represent these states as~$\cvect{v}'$ and~$\cvect{u}_k'$ with respect to some other measurement~$\mment{M}'$.  The no-disturbance postulate can then be used to relate the new representation to the old representation by completely coarse-graining measurement~$\mment{M}'$ and inserting measurement~$\mment{M}$ immediately afterwards.  If a state is represented by~$\cvect{v}'$ with respect to measurement~$\mment{M}'$, and~$\cmatrix{V}$ is the transformation matrix from~$\mment{M}'$ to~$\mment{M}$, then
\begin{equation}
v_i = \sum_j v'_{j} \, V_{ij} = (\cmatrix{V}\cvect{v}')_i,
\end{equation}
so that~$\cvect{v} = \cmatrix{V}\cvect{v}'$.  Since~$\cmatrix{V}$ is unitary, this can be inverted to give~$\cvect{v}' = \cmatrix{V}^\dagger\cvect{v}$.  Similarly~$\cvect{u}_k' = \cmatrix{V}^\dagger \cvect{u}_k$, which implies that the Hermitian operator,~$\cmatrix{N}'$, that represents~$\mment{N}$ with respect to~$\mment{M}'$, is given by~$\cmatrix{V}^\dagger\cmatrix{N}\cmatrix{V}$.

If measurement~$\mment{M}'$ is represented with respect to~$\mment{M}$ by Hermitian matrix~$\cmatrix{M}'$ with eigenvectors~$\cvect{w}_i$, then the transformation matrix~$
\cmatrix{V}$ has components~$V_{ij} = (\cvect{w}_j)_i$.

\subsection{Unitary Representation of Temporal Evolution}

Suppose that a system is prepared using measurement~$\mment{L}$ at time~$t$, and then undergoes measurement~$\mment{M}$ at time~$t'$.  Immediately prior to measurement~$\mment{M}$, the system is in state~$\cvect{v}$.   Suppose now that measurement~$\mment{M}$ is completely coarse-grained, and an additional measurement~$\mment{M}$ is performed at time~$t'' > t'$.   In this arrangement, the sequence~$\lseq{\outcome{\ell}{i}}{\widetilde{m}}{\outcome{m}{k}}$ can be decomposed as
\begin{equation}
\lseq{\outcome{\ell}{i}}{\widetilde{m}}{\outcome{m}{k}}  = \bigvee_j \,\seq{\outcome{\ell}{i}}{\outcome{m}{j}} \ser \seq{\outcome{m}{j}}{\outcome{m}{k}}.
\end{equation}
The temporal evolution of the system in interval~$[t', t'']$ is represented by the amplitudes~$U_{kj}$ of the sequences~$\seq{\outcome{m}{j}}{\outcome{m}{k}}$.  The amplitude sum and product rules accordingly imply that the amplitude~$\tilde{v}_k$ of sequence~$\lseq{\outcome{\ell}{i}}{\widetilde{m}}{\outcome{m}{k}}$ is
\begin{equation}
\tilde{v}_k =  \sum_{j} v_j U_{kj} = (\cmatrix{U}\cvect{v})_k.
\end{equation}
Therefore, the state of the system,~$\tilde{\cvect{v}}$, immediately prior to the last measurement is
\begin{equation}
\tilde{\cvect{v}}= \cmatrix{U}\cvect{v}.
\end{equation}
Now, the initial state~$\cvect{v}$ can be arbitrarily varied, but the states~$\cvect{v}$ and~$\tilde{\cvect{v}}$ are normalized.  Therefore,~$\cmatrix{U}$ is unitary.

Since temporal evolution from~$t'$ to~$t''$ is represented by~$\cmatrix{U}$, temporal evolution from~$t''$ to~$t'$ is represented by~$\cmatrix{U}^\dagger$.  Therefore, if we denote the temporal inverse of the sequence~$A$ as~$\invseq{A} $, then
\begin{equation} \label{eqn:temporal-inversion}
z(\invseq{A}) = z^*(A),
\end{equation}
to which we shall refer as the \emph{amplitude temporal inversion rule}.

\subsection{Composite Systems}

At time~$t$, system~$S_1$ is prepared by measurement~$\mment{L}_1$ with outcome~$\ell_1$, and~$S_2$ by measurement~$\mment{L}_2$ with outcome~$\ell_2$.   Suppose that these systems evolve without interacting with one another until time~$t'$, at which point they are measured, respectively, by~$\mment{M}_1$ and~$\mment{M}_2$, which respectively have $N_1, N_2$~possible outcomes,~$\outcome{m_1}{j}$ and~$\outcome{m_2}{k}$, with~$j \in \{1,\dots, N_1\}$ and~$k \in \{1, \dots, N_2\}$.  Then the components of the respective states,~$\cvect{v}'$ and~$\cvect{v}''$, of the systems immediately prior to~$t'$ are given by
\begin{equation}
v_j' = \amp\bigl(\seq{\ell_1}{\outcome{m_1}{j}}\bigr) \quad\quad\text{and}\quad\quad v_k'' = \amp\bigl(\seq{\ell_2}{\outcome{m_2}{k}}\bigr).
\end{equation}

Viewed as a single system,~$S$, the system is prepared by measurement~$\mment{L}$ with outcome~$\coutcome{\ell_1}{\ell_2}$, and subsequently undergoes measurement~$\mment{M}$ with~$N_1N_2$ possible outcomes~$\coutcome{\outcome{m_1}{j}}{\outcome{m_2}{k}}$.  With respect to~$\mment{M}$, the components of the state of the system,~$\cvect{v}$, immediately prior to~$\mment{M}$ is given by
\begin{equation}
v_{(i-1)N_2 + j} = \amp\bigl(\seq{\coutcome{\ell_1}{\ell_2}}{\coutcome{\outcome{m_1}{j}}{\outcome{m_2}{k}}} \bigr)
\end{equation}
But, by the composite systems rule, Eq.~\eqref{eqn:composite}, which we shall derive in Sec.~\ref{sec:composite-systems},
\begin{equation}
\amp\bigl(\seq{\coutcome{\ell_1}{\ell_2}}{\coutcome{\outcome{m_1}{j}}{\outcome{m_2}{k}}} \bigr) =  \amp\bigl(\seq{\ell_1}{\outcome{m_1}{j}}\bigr) \cdot \amp\bigl(\seq{\ell_2}{\outcome{m_2}{k}}\bigr),
\end{equation}
so that~$v_{(i-1)N_2 + j} = v_j' \, v_k''$.  Hence,
\begin{equation}
\cvect{v} = \cvect{v}' \otimes \cvect{v}''.
\end{equation}

\subsection{Summary}

In summary, we have derived the following:
\begin{enumerate}[(i)]
	\item If a system is prepared by some measurement~$\mment{L}$ at time~$t$, then its state at time~$t'$ immediately prior to measurement~$\mment{N}$ is given by vector~$\cvect{v}$ with respect to reference measurement~$\mment{M}$.

	\item Measurement~$\mment{N}$ is represented by Hermitian operator~$\cmatrix{N} = \sum_q a_q \cvect{u}_q \cvect{u}_q^\dagger$, where the~$\cvect{u}_q$, also  specified with respect to~$\mment{M}$, are the states prepared by~$\mment{N}$.  When performed on the system in state~$\cvect{v}$, the probability of the~$k$th outcome of~$\mment{N}$ is given by~$\Pr(\outcome{n}{k}\,|\,\cvect{v}) = |\cvect{u}_k^\dagger\cvect{v}|^2.$

	\item If the reference measurement is changed to~$\mment{M}'$, then the state~$\cvect{v}'$ of the system with respect to~$\mment{M}'$ is given by~$\cvect{v}' = \cmatrix{V}^\dagger\cvect{v}$, where~$\cmatrix{V}$ is the unitary transformation matrix from~$\mment{M}'$ to~$\mment{M}$.

	\item The state~$\cvect{v}$ evolves unitarily in the time between measurements.
	
	\item If the two subsystems of a composite system are in states~$\cvect{v}'$ and~$\cvect{v}''$, then the composite system is in state~$\cvect{v} = \cvect{v}' \otimes \cvect{v}''$.
	
	\end{enumerate}
	
Collectively, this set of statements is equivalent to von Neumann's postulates for finite-dimensional quantum systems.

%
%

\section{Composite Systems}
\label{sec:composite-systems}

\subsection{Composition Operator and its Symmetries}

Suppose that one physical system, denoted~$S_1$, undergoes an experiment involving measurements~$\mment{L}_1, \mment{M}_1, \mment{N}_1$ at successive times~$t_1, t_2, t_3$, while another system,~$S_2$, undergoes an experiment where the measurements~$\mment{L}_2, \mment{M}_2, \mment{N}_2$ at these same times.   The measurements on~$S_1$ yield the outcome sequence~$A = \lseq{\ell_1}{m_1}{n_1}$, while the measurements on~$S_2$ yield~$B = \lseq{\ell_2}{m_2}{n_2}$.  As described earlier, in Sec.~\ref{sec:experimental}, one can also describe the situation by saying that measurements~$\mment{L}, \mment{M}$ and~$\mment{N}$ are performed on the composite system~$S$, yielding the sequence~$C  = \lseq{\coutcome{\ell_1}{\ell_2}}{\coutcome{m_1}{m_2}}{\coutcome{n_1}{n_2}}$.    We now symbolize the relationship between~$A,B$ and~$C$ by defining a binary operator~$\comp$, the \emph{composition} operator, which here acts on~$A$ and~$B$ to generate the sequence
\begin{equation} \label{eqn:comp-operator}
C = A \comp B.
\end{equation}
Generally, the operator~$\comp$ combines any two sequences of the same length, each obtained from a different experiments on different physical systems where each measurement in one experiment occurs at the same time as one measurement in the other experiment.

\begin{figure*}
\includegraphics[width=6in]{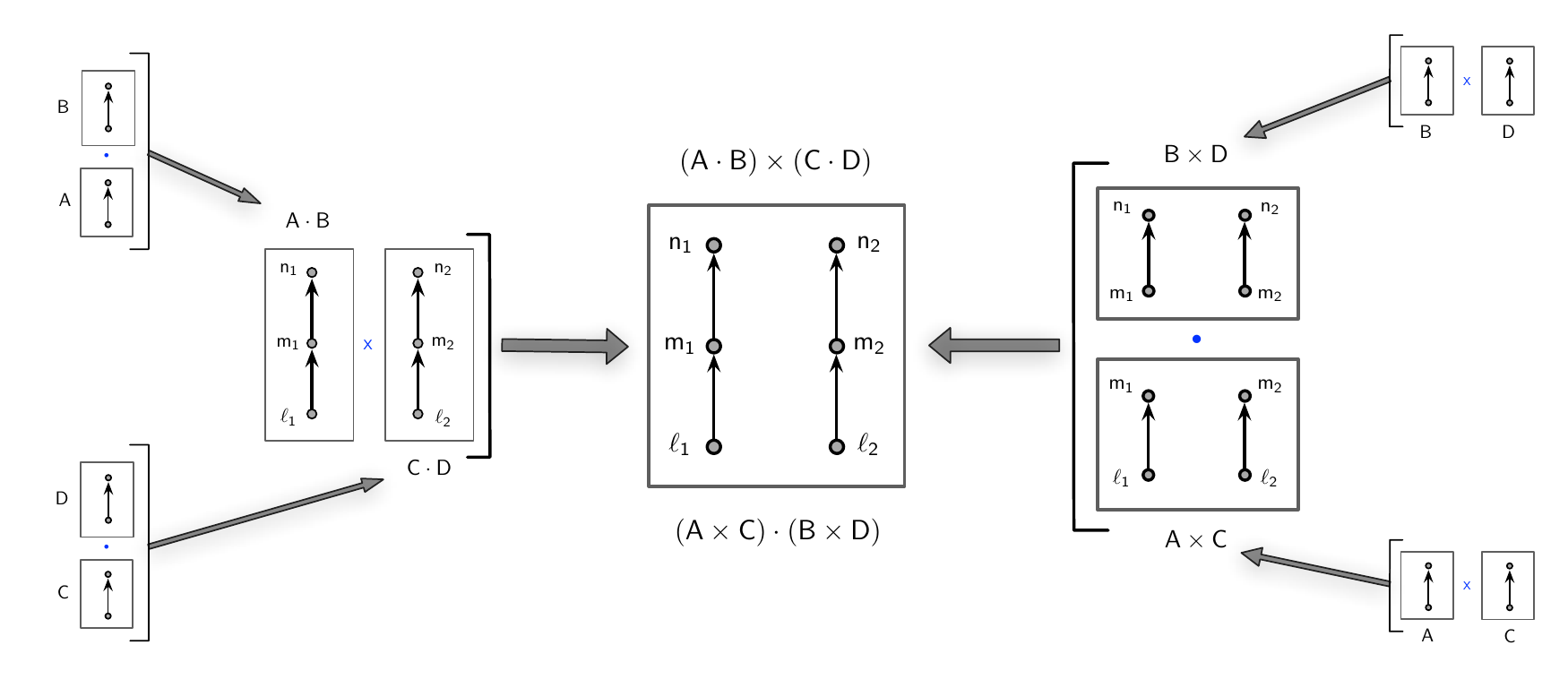}
\caption{\label{fig:comp-cross-multi} \emph{Illustration of the cross-multiplicativity property.}    The composite sequence~$\lseq{\coutcome{\ell_1}{\ell_2}}{\coutcome{m_1}{m_2}}{\coutcome{n_1}{n_2}}$ can be obtained by combining the sequences~$A = \seq{\ell_1}{m_1}, B = \seq{m_1}{n_1}, C = \seq{\ell_2}{m_2}$ and~$D = \seq{m_2}{n_2}$ in two different ways, as shown, yielding the cross-multiplicativity relation~$(A \ser B) \comp (C \ser D) = (A \comp C) \ser (B \comp D)$, which is Eq.~\eqref{eqn:comp-cross-multiplicativity}.
}
\end{figure*}

\begin{figure*}
\includegraphics[width=6in]{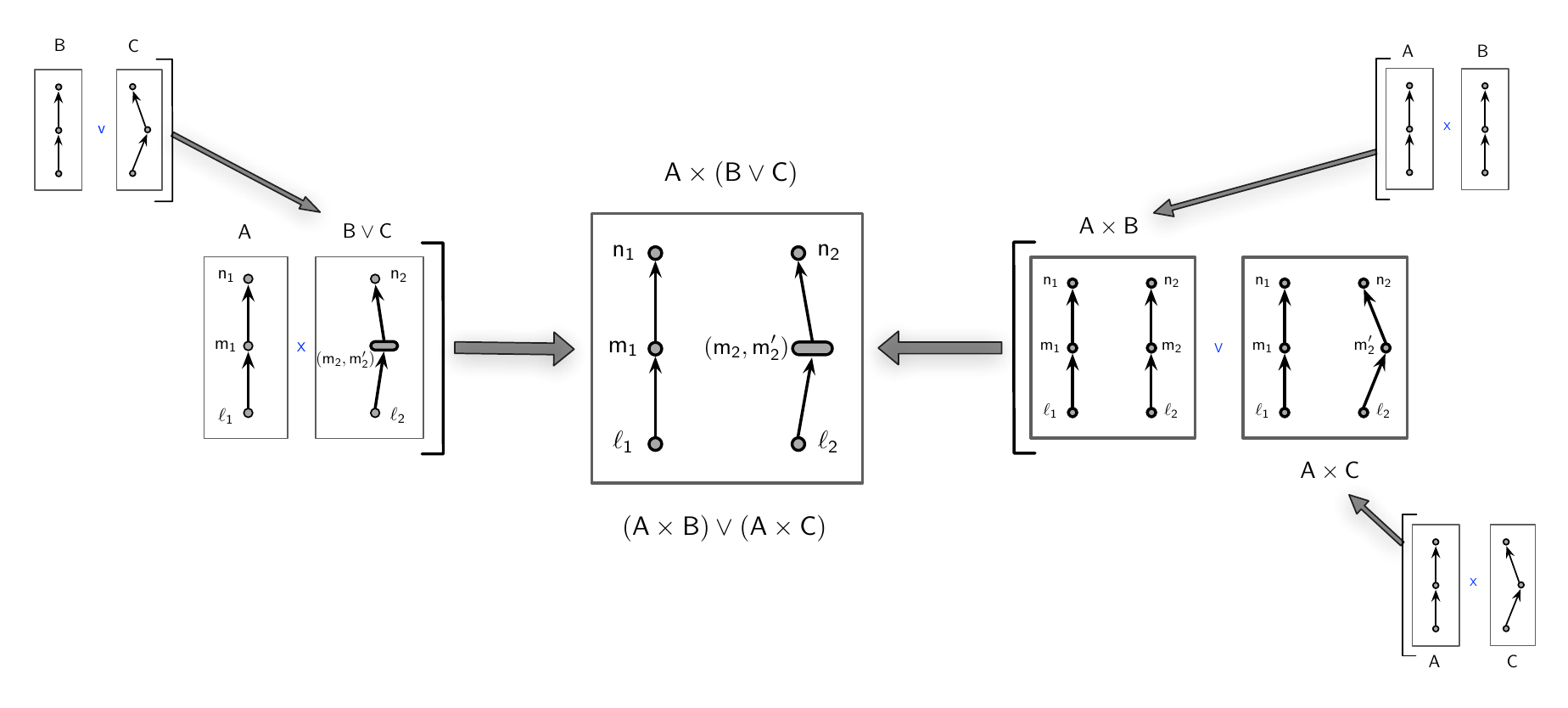}
\caption{\label{fig:comp-distributivity} \emph{Illustration of left-distributivity of~$\comp$ over~$\pll$.} The composite sequence~$\lseq{\bubble{\ell_1}{\ell_2}}{\coutcome{m_1}{\bubble{m_2}{m_2'}}}{\bubble{n_1}{n_2}}$ can be obtained by combining the sequences~$A = \lseq{\ell_1}{m_1}{n_1}$, $B = \lseq{\ell_2}{m_2}{n_2}$ and~$C = \lseq{\ell_2}{m_2'}{n_2}$ in two different ways, as shown, yielding the relation~$A \comp (B \pll C) = (A \comp B) \pll (A \comp C)$, which is Eq.~\eqref{eqn:comp-left-distributivity}. }
\end{figure*}

From the definition just given, it follows that~$\comp$ is associative.  To see this, consider the three sequences~$A = \seq{\ell_1}{m_1}$, $B = \seq{\ell_2}{m_2}$ and~$C = \seq{\ell_3}{m_3}$, obtained from three different experiments satisfying the condition just stated above.  We can then combine these to yield the sequence~$D = \seq{\coutcomel{\ell_1}{\ell_2}{\ell_3}}{\coutcomel{m_1}{m_2}{m_3}}$ in two different ways, either as~$A \comp (B \comp C)$ or as~$(A \comp B) \comp C$.  Hence,
\begin{align}
A \comp (B \comp C) &= (A \comp B) \comp C.  \label{eqn:comp-associativity}
\end{align}
Similar considerations show that~$\comp$ also satisfies the following symmetry relations involving the operators~$\pll$ and~$\ser$: 
\begin{align}
(A \ser B) \comp (C \ser D) &= (A \comp C) \ser (B \comp D) \label{eqn:comp-cross-multiplicativity} \\
A \comp (B \pll C) 	&= (A \comp B) \pll (A \comp C) \label{eqn:comp-left-distributivity} \\
(A \pll B) \comp C 	&= (A \comp C) \pll (B \comp C). \label{eqn:comp-right-distributivity} 
\end{align}
The cross-multiplicativity  and left-distributivity properties expressed in Eqs.~\eqref{eqn:comp-cross-multiplicativity} and~\eqref{eqn:comp-left-distributivity} are illustrated in Figs.~\ref{fig:comp-cross-multi} and~\ref{fig:comp-distributivity}, respectively.

%


\subsection{Composite Systems Rule}

If systems~$S_1$ and~$S_2$ are noninteracting, we postulate that, in Eq.~\eqref{eqn:comp-operator}, the amplitude,~$c$, of sequence~$C$ is determined by the amplitudes~$a, b$ of the sequences~$A, B$, so that
\begin{equation} \label{eqn:composition}
c = F(a, b),
\end{equation}
where~$F$ is some continuous complex-valued function to be determined.  This is the \emph{composition postulate}, given which Eqs.~\eqref{eqn:comp-associativity},~\eqref{eqn:comp-cross-multiplicativity}, \eqref{eqn:comp-left-distributivity} and~\eqref{eqn:comp-right-distributivity} respectively imply
\begin{align}
F(a, F(b, c)) 	&= F(F(a, b), c) \label{eqn:F-associativity}  					\\
F(ab, cd) 		&= F(a, c) \, F(b, d) \label{eqn:F-cross-multiplicativity}  	\\
F(a, b + c)		&= F(a, b) + F(a, c) \label{eqn:F-left-distributivity}			\\
F(a+b, c)		&= F(a, c) + F(b, c). \label{eqn:F-right-distributivity}
\end{align}
We can now solve these for~$F$.  Due to the cross-multiplicativity equation, Eq.~\eqref{eqn:F-cross-multiplicativity},
\begin{equation} \label{eqn:F-product}
F(u, v) = F(u\cdot 1, 1\cdot v) = F(u, 1) \, F(1, v).
\end{equation}
To determine form of~$F(u, 1)$, we use the right-distributivity and cross-multiplicativity equations, Eqs.~\eqref{eqn:F-right-distributivity} and~\eqref{eqn:F-cross-multiplicativity}, respectively, to obtain
\begin{subequations}
\begin{align*}
F(u_1 + u_2, 1) 	&= F(u_1, 1) + F(u_2, 1) \\
F(u_1 u_2 , 1) 		&= F(u_1, 1) \, F(u_2, 1).
\end{align*}
\end{subequations}
Writing~$f(z) = F(z, 1)$, these two equations can be written as a pair of functional equations,
\begin{subequations} \label{eqn:pair-functional-eqs}
\begin{align}
f(z_1 + z_2) 			&= 	f(z_1) + f(z_2) \label{eqn:pair-functional-eqs1}\\
f(z_1 z_2) 				&=	f(z_1) \, f(z_2), \label{eqn:pair-functional-eqs2}
\end{align}
\end{subequations}
whose continuous solutions in the domain~$|z| \leq 1$ are~$f(z) = z$, $f(z) = z^*$ or~$f(z) = 0$~(see Appendix~\ref{sec:functional}).  The zero solution implies~$F(u, v) = 0$ for all~$u, v$, and is therefore inadmissible.  Therefore,
\begin{subnumcases}{F(u, 1) =}
u \\
u^*.
\end{subnumcases}
To eliminate the possibility~$F(u, 1) = u^*$ we make use of the associativity equation, Eq.~\eqref{eqn:F-associativity}, which implies
\begin{equation} \label{eqn:associativity-example}
F(u, F(1, 1)) = F(F(u, 1), 1).
\end{equation}
Now, from the cross-multiplicativity equation, Eq.~\eqref{eqn:F-cross-multiplicativity},~$F(u\!\cdot\!1, \, v\!\cdot\!1) = F(u, v) \, F(1, 1)$, which implies~$F(1, 1)=1$ since the zero solution for~$F(u, v)$ is inadmissible.  Therefore, Eq.~\eqref{eqn:associativity-example} becomes
\begin{equation} \label{eqn:F-constraint}
F(u, 1) = F(F(u, 1), 1).
\end{equation}
But this is incompatible with~$F(u, 1) = u^*$ since~$u^* = F(u, 1) \neq F(F(u,1), 1) = F(u^*, 1) = u$.  We are therefore left with~$F(u, 1) = u$, which is compatible with Eq.~\eqref{eqn:F-constraint}.  A parallel argument establishes that~$F(1, v) = v$.  Therefore, from Eq.~\eqref{eqn:F-product},
\begin{equation} \label{eqn:composite}
F(u, v) = uv.
\end{equation}
This is the amplitude rule for distinguishable, noninteracting composite systems.  We refer to it as the \emph{composite system rule}.



\section{Derivation of Dirac's amplitude--action rule}
\label{sec:amplitude-action}

Consider a quantum system that is subject to position measurements at a successive times.  Suppose that the intervals between these successive times are sufficiently small that the resulting measurement sequence is well approximated by a continuous classical trajectory~(or simply `path') of the same system as treated according to the framework of classical physics.  Dirac's amplitude--action rule asserts that the amplitude associated with the measurement sequence is given by~$e^{iS/\hbar}$, where~$S$ is the classical action associated with the corresponding classical path.  We now derive the form of this rule up to~$\hbar$ from two elementary properties of the classical action, namely:
\begin{enumerate}
\item[] \emph{Additivity.}  If sequence~$C = A\ser B$, then~$S_C = S_A + S_B$, where~$S_X$ is the classical action of the path corresponding to sequence~$X$.
 
\item[] \emph{Inversion.} The action~$S_{\invseq{A}}$ associated with sequence~$\invseq{A}$ is~$-S_A$.
\end{enumerate}
Our assumption is that the amplitude,~$z(A)$, of sequence~$A$, with corresponding classical action is~$S_A$, is given by~$f(S_A)$, where~$f$ is a continuous, complex-valued function.  

\begin{figure}
\centering
\includegraphics[width=3.4in]{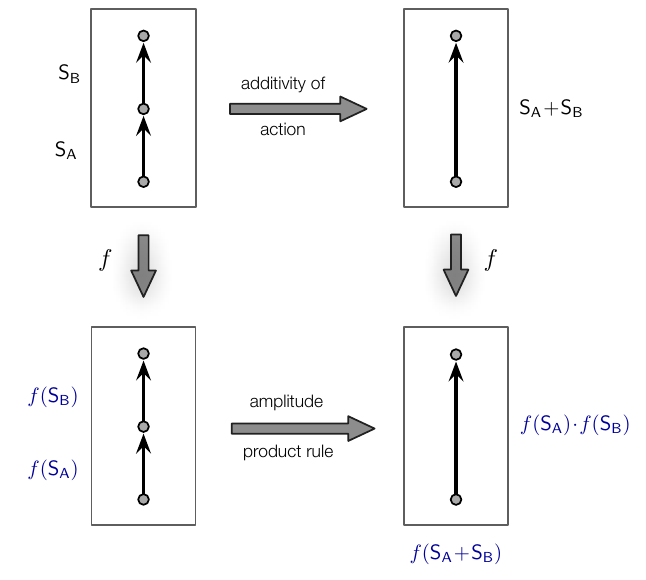}
\caption{\label{fig:amplitude-action} The amplitude of the path~$\sseq{C} = \sseq{A}\ser\sseq{B}$ can be obtained from the classical actions~$S_A, S_B$ of paths~$\sseq{A}, \sseq{B}$ in two different ways: (i)~obtain the action~$S_C = S_A + S_B$, whose corresponding amplitude is~$f(S_A + S_B)$; (ii)~use~$f$ to obtain the amplitudes~$f(A), f(B)$ and then compose these to obtain amplitude~$f(S_A)\cdot f(S_B)$.  Hence,~$f(S_A + S_B) = f(S_A)\cdot f(S_B)$.}
\end{figure}

The additivity and inversion properties of the classical action induce two functional equations in~$f$.  First, the amplitude~$z(C)$ of sequence~$C = A\ser B$ can be computed in two ways~(see Fig.~\ref{fig:amplitude-action}), %
either using the action additivity property,
\begin{equation*}
z(C) = f(S_C) = f(S_A + S_B),
\end{equation*}
or using the amplitude product rule,
\begin{equation*}
z(C) = z(A) \, z(B) = f(S_A) \, f(S_B),
\end{equation*}
so that
\begin{equation} \label{eqn:action-function-additivity}
f(x + y) = f(x) \, f(y).
\end{equation}

Second, the amplitude~$z(\invseq{A})$ of sequence~$\invseq{A}$ can be computed either using the amplitude temporal inversion rule, Eq.~\eqref{eqn:temporal-inversion},
\begin{equation*}
z(\invseq{A}) = z^*(A) = f^*(S_A),
\end{equation*}
or using the action inversion property,
\begin{equation*}
z(\invseq{A}) = f(S_{\invseq{A}}) = f(-S_A),
\end{equation*}
so that
\begin{equation} \label{eqn:action-function-inversion}
f^*(x) = f(-x).
\end{equation}

We can solve Eq.~\eqref{eqn:action-function-additivity} by writing~$f(x) = R(x)\,e^{i\Phi(x)}$, with~$R, \Phi$ real, to obtain, for integer~$n$,
\begin{align}
R(x + y) 		&= R(x) \, R(y)  \label{eqn:R-sum}\\
\Phi(x + y) 		&= \Phi(x) + \Phi(y) + 2\pi n.
\end{align}
The second equation can be transformed via~$\tilde{\Phi}(x) = \Phi(x) + 2\pi n$ to give
\begin{equation} \label{eqn:phi-sum}
\tilde{\Phi}(x + y) = \tilde{\Phi}(x) + \tilde{\Phi}(y).
\end{equation}
Equations~\eqref{eqn:R-sum} and~\eqref{eqn:phi-sum} are two of Cauchy's standard function equations, with general solutions~$R(x) = e^{\beta x}$ and~$\tilde{\Phi}(x) = \alpha x$, where~$\alpha, \beta \in\numberfield{R}$~\cite{Aczel-lectures-functional-equations}.  Hence, $\Phi(x) = \alpha x - 2\pi n$, and
\begin{equation}
f(x) = e^{\beta x} \, e^{i\alpha x}.
\end{equation}
But Eq.~\eqref{eqn:action-function-inversion} then implies that~$\beta = 0$.  Therefore,
\begin{equation}
z(A)  =  e^{i\alpha S_A},
\end{equation}
where the constant~$\alpha$ has dimensions of~$\hbar^{-1}$.  This is Dirac's amplitude--action rule up to~$\hbar$.


\section{Discussion}
\label{sec:discussion}

In this paper, we have shown that it is possible to systematically build Feynman's rules into a complete formulation of finite-dimensional quantum theory.  The key physical ingredient in this process has been the no-disturbance postulate, which expresses the singularly non-classical fact that a trivial measurement does not disturb the outcome probabilities of subsequent measurements on the system.  This postulate allows us to introduce the concept of the state of a system in a systematic way, and to prove the unitarity of temporal evolution and the hermiticity of measurement operators.   We have also derived the composite system rule and Dirac's amplitude--action rule, each from a single elementary and natural assumption, by making use of the fact that these assumptions must be consistent with Feynman's rules.

The work described here, in concert with our earlier derivation of Feynman's rules, constitutes a complete derivation of the finite-dimensional quantum formalism.   The derivation has a number of important implications for our understanding of quantum theory in addition to those mentioned in the Introduction.

First, most other attempts to derive the quantum formalism from physically-motivated postulates~(such as~\cite{Hardy01a, Hardy2013, Chiribella2011}) depend upon postulates~(such as purifiability~\cite{Chiribella2011} or local tomography~\cite{Hardy01a,Hardy2013}) that concern the behavior of composite systems in order to derive the quantum formalism for \emph{individual} systems.  This tends to suggest that the behavior of composite systems is in some sense fundamental to the structure of the quantum formalism.  However, in the present derivation, there is no such dependency.   Instead, we have shown that it is possible to derive the formalism for individual systems \emph{without} assumptions that overtly concern composite systems, and then to derive the rule for composite systems on the basis of a simple assumption merely by requiring consistency with the formalism for individual systems.  Therefore, the present derivation strongly implies that the behavior of composite systems is a secondary feature of quantum theory, not a primary one.

Second, one of the most remarkable features of Feynman's formulation of quantum theory is the absence of a state concept, and the absence of any distinction between dynamics, on the one hand, and the relationship between measurements, on the other.  We have shown here that these features \emph{can} be recovered, but at the cost of an additional physical postulate which has non-trivial physical content.  





Finally, we have shown that Dirac's amplitude--action rule follows from elementary properties of the classical action via the simple assumption that the amplitude of a sequence is determined by the corresponding action.   In contrast with Dirac's argument, our approach does not depend on the particular form of the classical Lagrangian or on the existence or form of Lagrange's equations of motion, but only on two elementary properties~(additivity, inversion) of the action.   Hence, we have shown that Dirac's rule has a very general validity, and arises as soon as one attempts to establish a quantitative connection between the notion of action in the Lagrangian formulation of classical mechanics, and the notion of amplitude in Feynman's formulation of quantum theory.

We conclude with two open questions.  First, is the no-disturbance postulate related in any way with other informational ideas that have been proposed, such as in Refs.~\cite{Brukner99,Brukner-Zeilinger2009,Chiribella2011}?  Second, is there a direct, general path from Dirac's amplitude--action rule to the unitary form,~$\exp\bigl(-i\hat{H}\Delta t / \hbar\bigr)$, of the temporal evolution operator?

\begin{acknowledgments}
This publication was, in part, made possible through the support of a grant from the John Templeton Foundation.
\end{acknowledgments}


\clearpage
\begin{widetext}
\begin{appendix}

\section{Solution of a pair of functional equations.}
\label{sec:functional}

We solve Eqs.~\eqref{eqn:pair-functional-eqs1} and~\eqref{eqn:pair-functional-eqs2} with the aid of one of Cauchy's standard functional equations,
\begin{equation} \label{eqn:Cauchy}
h(x_1 + x_2) = h(x_1) + h(x_2),
\end{equation}
where~$h$ is a real function and~$x_1, x_2 \in \numberfield{R}$.  Its continuous solution is~$h(x) = ax$ with~$a \in\numberfield{R}$~\cite{Aczel-lectures-functional-equations}.

Setting~$z_1 + z_2 = x + iy$, with~$x, y \in\numberfield{R}$, in Eq.~\eqref{eqn:pair-functional-eqs1} gives
\begin{equation*} \label{eqn:f-decomp}
f(x + iy) = f(x) + f(iy).
\end{equation*}
Applying Eq.~\eqref{eqn:pair-functional-eqs1} again on~$f(x_1 + x_2)$ and~$f(iy_1 + iy_2)$ then implies 
\begin{align*}
f(x_1 + x_2) &= f(x_1) + f(x_2)  \\
f\left(i y_1 + i y_2 \right) &= f(iy_1) + f(iy_2).
\end{align*}
The real and imaginary parts of both of these equations each have the form of Eq.~\eqref{eqn:Cauchy}, and therefore have solutions
\begin{equation*}
f(x) = \alpha x  \quad\quad\text{and}\quad\quad f(iy) = \beta y
\end{equation*}
with~$\alpha, \beta \in \numberfield{C}$, so that
\begin{equation} \label{eqn:f-form-after-additivity}
f(x+iy) = \alpha x + \beta y.
\end{equation}

From Eq.~\eqref{eqn:pair-functional-eqs2},
\begin{equation*}
f(1 \cdot 1) = f(1) f(1)  \quad\quad\text{and}\quad\quad f(i \cdot i) = f(i) f(i),
\end{equation*}
which, due to Eq.~\eqref{eqn:f-form-after-additivity}, imply
\begin{equation*}
\alpha = \alpha^2   \quad\quad\text{and}\quad\quad  {-\alpha} = \beta^2.
\end{equation*}
These have solutions~$(\alpha, \beta) = (0,0), (1,  i)$ and~$(1, -i)$, which correspond to~$f(z) = 0$, $f(z) = z$ and~$f(z) = z^*$.

\end{appendix}
\clearpage
\end{widetext}

\end{document}